\documentclass{Dekker}
\usepackage{amsmath}
\usepackage{amssymb}
\begin{document}
\setcounter{page}{1}
\jname{TRANSPORT THEORY AND STATISTICAL PHYSICS}
\jvol{ }
\jissue{ }
\jyear{}
\webslug{www.dekker.com}
\cpright{Marcel Dekker, Inc.}
\title[CAUCHY INTEGRAL EQUATIONS. III]{RADIATIVE TRANSFER IN PLANE-PARALLEL MEDIA AND CAUCHY INTEGRAL EQUATIONS III. THE FINITE CASE}
\author[RUTILY, CHEVALLIER AND BERGEAT]{B. Rutily, L. Chevallier, J. Bergeat
\affiliation{Centre de Recherche Astronomique de Lyon (UMR 5574 du CNRS),
Observatoire de Lyon,\\
9, Avenue Charles Andr\'e,
69561 Saint-Genis-Laval Cedex, France \\ \ \\ Submitted January 2003}}

\abstract{We come back to the Cauchy integral equations occurring in radiative transfer problems posed in finite, plane-parallel media with light scattering taken as mono\-chromatic and isotropic. Their solution is calculated following the classical scheme where a Cauchy integral equation is reduced to a couple of Fredholm integral equations. It is expressed in terms of two auxiliary functions $\zeta_+$ and $\zeta_-$ we introduce in this paper. These functions show remarkable analytical properties in the complex plane. They satisfy a simple algebraic relation which generalizes the factorization relation of semi-infinite media. They are regular in the domain of the Fredholm integral equations they satisfy, and thus can be computed accurately. As an illustration, the $X$- and $Y$-functions are calculated in the whole complex plane, together with the extension in this plane of the so-called Sobouti's functions}

\keywords{Radiative transfer equation; Finite, plane-parallel medium; Isotropic scattering; Cauchy integral equations}

\maketitle

\section{\label{Sec1} INTRODUCTION}
In Refs. [1] and [2] (hereafter I and II), we revisited the Cauchy integral equations arising in the simplest radiative transfer problems posed in plane-parallel geometry, that is in homogeneous and stationary media with monochromatic and isotropic light scattering. The literature on this subject essentially dates back to the sixties.$^{[3-8]}$ In the present paper, we deal with singular integral equations of the form
\begin{eqnarray}
\lefteqn{T(a,z) X_{0}(a,b,z)-\frac{a}{2}z\int_{0}^{1}X_{0}(a,b,v)\frac{dv}{v-z}} \nonumber \\&&\; +\exp(-b/z)\frac{a}{2}z\int_{0}^{1}X_{0}(a,b,-v)\exp(-b/v)\frac{dv}{v+z}=c_{0}(z),
\end{eqnarray}
which are encountered when solving transfer problems in finite slabs$^{[9]}$ The parameters $a$ and $b$ are respectively the volumic albedo of the medium and its optical thickness at a fixed frequency ($0<a<1$ and $0<b\leq+\infty$). The dispersion function $T(a,z)$ is defined in $\mathbb{C}\setminus \lbrace-1,+1\rbrace$ by
\begin{equation}
T(a,z)=1+\frac{a}{2}z\int_{-1}^{+ 1}\frac{dv}{v-z}. 
\end{equation}

We specify that the integrals in Eqs. (1) and (2) are Cauchy principal values when the variable $z$ is in the range $]-1,+1[$, with  no change in notation: as in I and II, the same symbol is used for a sectionally analytic function outside and on its cut.

The $c_0$-function on the right-hand side of Eq. (1) is defined and analytic in $\mathbb{C}^* = \mathbb{C}\setminus\lbrace0\rbrace$, including infinity. As a matter of fact, $c_0(1/z)$ is the Laplace transform, over the finite interval $[0,b]$, of the source function $S_0$ for the direct field within the slab, viz.$^{[9]}$
\begin{equation}
c_{0}(z)=\int_{0}^{b}S_{0}(\tau)\exp(-\tau/z)d\tau. 
\end{equation}
As a result, this source term is bounded on the right at $z = 0$, which means that the limit of $c_0(z)$ exists and is finite when $z$ tends to 0 in the (stict) right complex half-plane. This remains true for the function $z\rightarrow c_0(-z)\exp(-b/z)$, as can be seen by replacing $S_0(\tau)$ by $S_0(b-\tau)$ in the integrand of Eq. (3). Whence
\begin{equation}
\lim_{z\rightarrow 0\,,\, \Re(z)>0}c_0(z)<+\infty\;,\;\lim_{z\rightarrow 0\,,\, \Re(z)>0}[c_0(-z)\exp(-b/z)]<+\infty,
\end{equation}
where $\Re(z)$ denotes the real part of the complex number $z$. We note that the possibility of $c_0$ defined and analytic at $z=0$ is not ruled out. It happens when $S_0(\tau)=\delta(\tau)$, the Dirac distribution at the origin. We have then $c_0(z) = 1$ and $c_0(-z)\exp(-b/z) = \exp(-b/z)$. These expressions are exchanged when $S_0(\tau)=\delta(b-\tau)$, i.e., $c_0(z)=\exp(-b/z)$ and $c_0(-z)\exp(-b/z) = 1$. If $S_0$ is a function or the Dirac distribution at $\tau\in]0,b[$, $c_0(z)$ and $c_0(-z)\exp(-b/z)$ are not defined at $z=0$ and the limits in (4) are zero.

The unknown function $X_0$ depends on the given source term $c_0$ and the parameters $a$ and $b$. This dependence is made explicit since the 0-subscript is intended to remind of the link with $c_0$. This subscript is removed when dealing with the particular free term $c_0 = 1$, the associated $X$-function being the well-known function of finite plane-parallel media. The function $X_0$ can be assumed analytic everywhere in $\mathbb{C}^*$, since $X_0(a,b,1/z)$ is the finite Laplace transform of the source function $S$ for the diffuse field within the slab.$^{[9]}$

The second integral term in Eq. (1) vanishes in a half-space ($b=+\infty$) when $\Re(z)>0$. This equation reduces to the simpler form
\begin{equation}
T(a,z)f(z)-\frac{a}{2}z\int_{0}^{1}f(v)\frac{dv}{v-z}=c(z), 
\end{equation}
where $c(z)=c_0(z)$. Thus it is of Cauchy type on $[0,1]$. We studied this kind of equation in I, and gave in II its unique solution analytic in the right complex half-plane. From now and subsequently, we refer to the equations of I (resp. II) by their original number preceded by the roman numeral I (resp. II). The solution (II, 50) to Eq. (5) was written for a free term $c$ verifying the three following properties:

\noindent
({\it i}) it is defined in $\mathbb{C}\setminus\lbrace-1\rbrace$, except possibly at $z=0$,

\noindent
({\it ii}) it is analytic in the half-plane $\Re(z)\geq0$ (including infinity) except possibly at $z=0$,

\noindent
({\it iii}) it is bounded on the right at $z=0$, i.e., the limit of $c(z)$ when $z\rightarrow0$ with $\Re(z)>0$ exists and is
finite.

\noindent
We have expressed this solution in terms of the particular solution $H = H(a,z)$, which corresponds to the free term $c(z) = 1$. This is just the well-known $H$-function for semi-infinite media.

In a finite space ($b<+\infty$), the two integral terms in Eq. (1) are kept, so that if this equation is written in the form (5), the free term $c(z)$ depends on the unknown function $X_0$. This substantial complication prevents us from solving Eq. (1) explicitly, as in the semi-infinite case.

The remainder of this article may be summarized as follows: in Sec. \ref{Sec2}, we use the classical transformation from Eq. (1), which is of Cauchy type over the line segment $[-1,+1]$, to two Cauchy integral equations over $[0,1]$. Their solution is expressed in terms of two auxiliary functions depending on the free term $c_0$. In Sec. \ref{Sec3}, we mainly deal with the study of the auxiliary functions associated to the particular free term $c_0 = 1$, which we denote $\zeta_+$ and $\zeta_-$. It is shown that these functions have remarkable analytical properties in the complex plane. In particular they satisfy a simple algebraic relation which generalizes the factorization relation for semi-infinite media. These functions allow the calculation of the $X$- and $Y$-functions associated to finite, plane-parallel media, together with the extension to the complex plane of the so-called Sobouti's functions. We come back to the general case in Sec. \ref{Sec4}, where it is shown that the general solution to Eq. (1) can be expressed by means of the functions $\zeta_+$ and $\zeta_-$ we studied in Sec. \ref{Sec3}. The latter are thus the basic functions for the solution of Eq. (1) with any free term $c_0$. Finally, an algorithm for the computation of $\zeta_+$ and $\zeta_-$ is proposed in Sec. \ref{Sec5}. It is based on the resolution of two uncoupled Fredholm integral equations with regular kernels.

\section{\label{Sec2} PRINCIPLE OF THE SOLUTION OF EQUATION (1)}
Due to its second integral term, Eq. (1) is of Cauchy type over $[-1,+1]$ while it was so over $[0,1]$ in the semi-infinite case. We are thus confronted with a first difficulty: the solution to Eq. (1) can't be H\"{o}lder continuous in $[-1,+1]$ since it diverges at zero. This very fact prevents us from applying directly the theory of Cauchy integral equations. This difficulty is overcome by replacing the solution to Eq. (1) over $[-1,+1]$ by the one over $[0,1]$ to a system of two singular integral equations satisfied by the functions $X_0(a,b,z)$ and $Y_0(a,b,z) = X_0(a,b,-z)\exp(-b/z)$. This classical approach was followed by Busbridge$^{[3]}$ and Mullikin et al.$^{[4-7]}$ among others.

The functions $X_0$ and $Y_0$ are related for any $z\in\mathbb{C}^*$ by
\begin{eqnarray}
X_0(a,b,z)&=&Y_0(a,b,-z)\exp(-b/z),\\ 
Y_0(a,b,z)&=&X_0(a,b,-z)\exp(-b/z), 
\end{eqnarray}
and we have for $z\in\mathbb{C}\setminus\lbrace\pm1,0\rbrace$
\begin{eqnarray}
\lefteqn{T(a,z)X_{0}(a,b,z)-\frac{a}{2}z\int_{0}^{1}X_{0}(a,b,v)\frac{dv}{v-z}}\nonumber\\
&&\qquad\qquad\qquad+\exp(-b/z)\frac{a}{2}z\int_{0}^{1}Y_{0}(a,b,v)\frac{dv}{v+z}=c_{0}(z), 
\end{eqnarray}
\begin{eqnarray}
\lefteqn{T(a,z)Y_{0}(a,b,z)-\frac{a}{2}z\int_{0}^{1}Y_{0}(a,b,v)\frac{dv}{v-z}}\nonumber\\
&&+\exp(-b/z)\frac{a}{2}z\int_{0}^{1}X_{0}(a,b,v)\frac{dv}{v+z}=c_{0}(-z)\exp(-b/z).
\end{eqnarray}
The second equation is deduced from the first one by changing $z$ into $-z$ in it, then multiplying the resulting equation by $\exp(-b/z)$.

The coupling between Eqs. (8) and (9) can be removed by substracting and adding them. We obtain
\begin{eqnarray}
\lefteqn{T(a,z)(X_{0}\mp Y_{0})(a,b,z)}\nonumber\\&&\qquad\qquad-\frac{a}{2}z\int_{0}^{1}(X_{0}\mp Y_{0})(a,b,v)\frac{dv}{v-z}=c_{0,\pm}(a,b,z),
\end{eqnarray}
where, for any $z\in\mathbb{C}\setminus\lbrace-1,0\rbrace$
\begin{eqnarray}
\lefteqn{c_{0,\pm}(a,b,z)=}\nonumber\\
&&c_0(z)\mp[c_{0}(-z)-\frac{a}{2}z\!\int_{0}^{1}\!(X_{0}\!\mp\! Y_{0})(a,b,v)\frac{dv}{v\!+\!z}]\exp(-b/z). 
\end{eqnarray}

This expression shows that the limits   
\begin{equation}
c_{0,\pm}(a,b,+0)=\lim_{z\rightarrow0,\Re(z)>0}c_{0,\pm}(a,b,z)
\end{equation}
are given by
\begin{equation} 
c_{0,\pm}(a,b,+0)=\lim_{z\rightarrow0,\Re(z)>0}[c_{0}(z)\mp c_{0}(-z)\exp(-b/z)] <+\infty,
\end{equation}
the inequality being due to (4).

Equation (10) is a Cauchy integral equation of the form (5) for the functions $X_0\mp Y_0$. Since its free term $c_{0,\pm}(a,b,z)$ satisfies the properties ({\it i})-({\it iii}) as given in the introduction, its unique solution analytic in the right complex half-plane is given by Eq. (II, 50), viz.
\begin{eqnarray}
\lefteqn{(X_{0}\mp Y_{0})(a,b,z)=}\nonumber\\
&&H(a,z)\left\lbrace \frac{1}{2}c_{0,\pm}(a,b,+0)+{\rm Y}[-\Re(z)]H(a,-z)c_{0,\pm}(a,b,z)\right.\nonumber\\&&\left.\qquad\qquad+\frac{z}{2i\pi}\!\int_{-i\infty}^{+i\infty}\!H(a,1/z')c_{0,\pm}(a,b,-1/z')\frac{dz'}{1\!+\!zz'}\right\rbrace. 
\end{eqnarray}
Here ${\rm Y}$ is the unit step (or Heaviside) function extended by 1/2 at 0: ${\rm Y}(x)$ is 0 for $x<0$, 1/2 for $x=0$ and +1 for $x>0$. The integral is a principal value at infinity. This relation is valid on the imaginary axis (denoted hereafter $i\mathbb{R}$), provided that the value $\rm Y(0)=1/2$ is adopted and the integral is taken in the sense of the Cauchy principal value.

We introduce the functions
\begin{equation}
u_{0,\pm}(a,b,z)=(X_0\mp Y_0)(a,b,z),
\end{equation}
\begin{equation}
v_{0,\pm}(a,b,z)=c_0(-z)-\frac{a}{2}z\int_{0}^{1}(X_0\mp Y_0)(a,b,v)\frac{dv}{v+z}.
\end{equation}

The functions $u_{0,\pm}$ are defined on $\mathbb{C}^*$ where they are analytic. They diverge on the right-hand side of 0. The functions $v_{0, \pm}$ are defined in $\mathbb{C}\setminus\lbrace-1,0\rbrace$. They are sectionally analytic over $\mathbb{C}\setminus[-1,0]$, the integral being a Cauchy principal value for $z\in]-1,0[$. They diverge at -1 due to the integral term, and possibly on the right side of 0 due to the $c_0(-z)$-term. The divergence at -1 results from the behavior of the functions $(X_0\mp Y_0)(a,b,v)$ on the left side of +1: they don't vanish there (see Ref. [10], p. 42).

It follows from Eqs. (6)-(7), (15)-(16) and (10)-(11) that the functions $u_{0,\pm}$ and $v_{0,\pm}$ are related by
\begin{eqnarray}
u_{0,\pm}(a,b,z)&=&\mp u_{0,\pm}(a,b,-z)\exp(-b/z)\,,\\
v_{0,\pm}(a,b,z)&=&c_0(-z)-\frac{a}{2}z\int_0^1u_{0,\pm}(a,b,v)\frac {dv}{v+z},\\
T(a,z)u_{0,\pm}(a,b,z)&=&v_{0,\pm}(a,b,\!-z)\mp v_{0,\pm}(a,b,z)\exp(\!-b/z),
\end{eqnarray}
and Eqs. (14) and (11) become
\begin{eqnarray}
\lefteqn{u_{0,\pm}(a,b,z)=}\nonumber\\
&&H(a,z)\left\lbrace\frac{1}{2}c_{0,\pm}(a,b,+0)+{\rm Y}[-\Re(z)]H(a,-z)c_{0,\pm}(a,b,z)\right.\nonumber\\
&&\qquad\qquad\left.+\frac{z}{2i\pi}\!\int_{-i\infty}^{+i\infty}\!H(a,1/z')c_{0,\pm}(a,b,-1/z')\frac{dz'}{1\!+\!zz'}\right\rbrace, 
\end{eqnarray}
\begin{equation}
c_{0,\pm}(a,b,z)=c_{0}(z)\mp v_{0,\pm}(a,b,z)\exp(-b/z). 
\end{equation}

Now multiply both sides of Eq. (20) by $T(a,z)$ in order to eliminate $u_{0,\pm}(a,b,z)$ with the help of Eq. (19), and substitute the expression (21) of $c_{0,\pm}$ in the right-hand side of Eq. (20). Using the factorization relation (II, 27) and changing $z$ into $-z$, one obtains
\begin{eqnarray}
\lefteqn{H(a,z)v_{0,\pm}(a,b,z)=\frac{1}{2}c_{0,\pm}(a,b,+0)+{\rm Y}[\Re(z)]H(a,z)c_{0}(-z)}\nonumber\\
&&\qquad\qquad\qquad\qquad\pm{\rm Y}[-\Re(z)]H(a,z)v_{0,\pm}(a,b,-z)\exp(b/z)\nonumber\\
&&\qquad\qquad\qquad\qquad-\frac{z}{2i\pi}\int_ {-i\infty}^{+i\infty}H(a,-1/z')\\
&&\qquad\qquad\qquad\qquad\times\!\!\left[c_{0}(1\!/\!z')\!\mp\! v_{0,\pm}(a,\!b,\!1/\!z')\exp(\!-bz')\right]\!\frac{dz'}{1\!\!+\!\!zz'}.\nonumber 
\end{eqnarray}

We introduce the functions $\zeta_{0,\pm}=\zeta_{0,\pm}(a,b,z)$ as defined in $\mathbb{C}\!\setminus\! i\mathbb{R}$ by
\begin{eqnarray}
\zeta_{0,\pm}(a,b,z)&=&H(a,z)v_{0,\pm}(a,b,z)\nonumber\\
&\mp&{\rm Y}[-\Re(z)]H(a,z)v_{0,\pm}(a,b,-z)\exp(b/z).
\end{eqnarray}
It follows from Eq. (22) that they satisfy the integral equations
\begin{eqnarray}
\zeta_{0,\pm}(a,b,z)&=&\frac{1}{2}c_{0,\pm}(a,b,+0)+{\rm Y}[\Re(z)]H(a,z)c_{0}(-z)\nonumber\\
&-&\frac{z}{2i\pi}\int_{-i\infty}^{+i\infty}H(a,-1/z')\\
&&\times[c_0(1\!/\!z')\mp\frac{\exp(-bz')}{H(a,\!1\!/\!z')}(\zeta_{0,\pm})^+(a,b,1\!/\!z')]\frac{dz'}{1\!+\!zz'},\nonumber
\end{eqnarray}
since $v_{0,\pm}(a,b,1/z')$ can be replaced by $(\zeta_{0,\pm})^+(a,b,1/z')/H(a,1/z')$ in the integral of Eq. (22), due to Eq. (23). The functions $(\zeta_{0,\pm})^+$ are the limit of the functions $\zeta_{0,\pm}$ on the right-hand side of the imaginary axis. Relation (24) shows that the latter functions are sectionally analytic in the complex plane cut along the imaginary axis, with limits $(\zeta_{0,\pm})^+$ and $(\zeta_{0,\pm})^-$ on the right- and left-hand sides of this axis respectively. We could extend the definition of $\zeta_{0,\pm}$ to the imaginary axis by calculating the integral in Eq. (24) in the sense of the Cauchy principal value; however, such an extension does not seem useful in the context of this article. We note in addition that the integral in Eq. (24) is a principal value at infinity. It can be transformed into an integral over $[0,1]$ using the residue theorem, which implies the existence and uniqueness of the functions $\zeta_{0,\pm}$ as defined by Eq. (24). Here we dispense with the details of this demonstration.

Replacing $v_{0,\pm}(a,b,-z)$ by $\zeta_{0,\pm}(a,b,-z)/H(a,-z)\; [\Re(z)<0]$ on the right-hand side of Eq. (23), we derive the following solution to Eq. (22) in $\mathbb{C}\setminus \lbrace i\mathbb{R}\cup\lbrace-1\rbrace\rbrace$:
\begin{equation}
v_{0,\pm}(a,b,z)=\frac{\zeta_{0,\pm}(a,b,z)}{H(a,z)}\pm{\rm Y}[-\Re(z)]\frac{\zeta_{0,\pm}(a,b,-z)}{H(a,-z)}\exp(b/z).
\end{equation}

Hence $u_{0,\pm}$ in $\mathbb{C}\setminus i\mathbb{R}$ using Eq. (19) and the factorization relation (II, 27)
\begin{eqnarray}
\lefteqn{u_{0,\pm}(a,b,z)={\rm Y}[\Re(z)]H(a,z)\zeta_{0,\pm}(a,b,-z)}\nonumber\\
&&\qquad\qquad\quad\mp{\rm Y}[-\Re(z)]H(a,-z)\zeta_{0,\pm}(a,b,z)\exp(-b/z). 
\end{eqnarray}

These expressions can be extended to the imaginary axis by letting $\Re(z)\to0^{\pm}$ in them and taking into account the continuity of the functions $u_{0,\pm}$ and $v_{0,\pm}$ on the $i\mathbb{R}$-axis. We derive their restriction on this axis in terms of the limits of the functions $\zeta_{0,\pm}$ on both sides of the axis. Hence for any $z_0 \in i\mathbb{R}^*$
\begin{eqnarray}
v_{0,\pm}(a,b,z_0)&=&\frac{(\zeta_{0,\pm})^+(a,b,z_0)}{H(a,z_0)}\,,\\&=&\frac{(\zeta_{0,\pm})^-\!(a,\!b,\!z_0)}{H(a,z_0)}\pm \frac{(\zeta_{0,\pm})^+\!(a,\!b,\!-z_0)}{H(a,-z_0)}\exp(b/\!z_0), 
\end{eqnarray}
\begin{eqnarray}
u_{0,\pm}(a,b,z_0)&=&H(a,z_0)(\zeta_{0,\pm})^-(a,b,-z_0)\,,\\&=&\mp H(a,-z_0)(\zeta_{0,\pm})^-(a,b,z_0)\exp(-b/z_0). 
\end{eqnarray}

We may "invert" these relations by writing on $\mathbb{C}\setminus i\mathbb{R}$
\begin{eqnarray}
\zeta_{0,\pm}(a,b,z)&=&{\rm Y}[\Re(z)]H(a,z)v_{0,\pm}(a,b,z)\nonumber\\&+&{\rm Y}[-\Re(z)]\frac{u_{0,\pm}(a,b,-z)}{H(a,-z)},
\end{eqnarray}
which shows that the functions $\zeta_{0,\pm}$ give information about $v_{0,\pm}$ from their values in the right complex half-plane, and about $u_{0,\pm}$ from their values in the left one.

The solution $X_0$ to Eq. (1) and related functions appearing in the left-hand side of this equation are easily deduced from the functions $u_{0,\pm}$ and $v_{0,\pm}$, since we have from Eqs. (15)-(16)
\begin{eqnarray}
X_0(a,b,z)&=&\frac{1}{2}(u_{0,+}+u_{0,-})(a,b,z)\qquad(z\in\mathbb{C}^*),\\ 
Y_0(a,b,z)&=&-\frac{1}{2}(u_{0,+}-u_{0,-})(a,b,z)\qquad(z\in\mathbb{C}^*), 
\end{eqnarray}
\begin{eqnarray}
c_0(-z)-\frac{a}{2}z\int_0^1X_{0}(a,b,v)\frac{dv}{v+z}=\nonumber\\\frac{1}{2}(v_{0,+}+v_{0,-})(a,b,z)&\quad(z\in\mathbb{C}\setminus\lbrace-1,0\rbrace),
\end{eqnarray}
\begin{equation}
\frac{a}{2}z\!\int_0^1\!Y_{0}(a,b,v)\frac{dv}{v\!+\!z}=\frac{1}{2}(v_{0,+}\!-v_{0,-})(a,b,z)\;(z\in\mathbb{C}\setminus\lbrace-1\rbrace). 
\end{equation}

Finally, the only thing we have to do is to solve the problem (24), since its solution $\zeta_{0,\pm}$ leads to the functions $u_{0,\pm}$ and $v_{0,\pm}$ via Eqs. (25)-(30), then to the solution to Eq. (1) via Eqs. (32)-(35). In the next section, we first solve Eq. (24) for the particular free term $c_0(z)=1$.

\section{\label{Sec3}THE $c_0=1$ CASE: CALCULATION OF THE $X$- AND $Y$-FUNCTIONS} 
We treat the case $c_0 = 1$ in this section. The associated functions will be denoted by the same symbol as in the general case, just removing the 0-subscript. Equation (24) simplifies for $c_0 = 1$, since we have $c_{\pm}(a,b,+0)=1$ from Eq. (12) and
\begin{equation}
\frac{z}{2i\pi}\int_{-i\infty}^{+i\infty}H(a,-1/z')\frac{dz'}{1+zz'}=-\frac{1}{2}+{\rm Y}[\Re(z)]H(a,z) 
\end{equation}
owing to the residue theorem applied to an obvious contour in the left complex half-plane. The functions $\zeta_{\pm}$ are thus defined on $\mathbb{C}\!\setminus \!i\mathbb{R}^*$ by the integral equations
\begin{eqnarray}
\lefteqn{\zeta_{\pm}(a,b,z)=1\pm \frac{z}{2i\pi}\int_{-i\infty}^{+i\infty}\frac{H(a,-1/z')}{H(a,1/z')}}\nonumber\\
&&\qquad\qquad\qquad\qquad\quad\times\exp(-bz')(\zeta_{\pm})^+(a,b,1/z')\frac{dz'}{1\!+\!zz'}.
\end{eqnarray}

These equations have a sense at $z=0$ and yield $\zeta_{\pm}(a,b,0)=1$, since the identity (36) is valid at $z=0$ [${\rm Y}(0)=1/2$]. In Sec. \ref{Sec5}, it will be seen that Eqs. (37) can be transformed into two uncoupled Fredholm integral equations over $[0,1]$. A detailed study of these equations will be published in a forthcoming paper. It can be shown that they admit a unique solution in their domain, which implies the existence and uniqueness of the functions $\zeta_{\pm}$ as defined on $\mathbb{C}\setminus i\mathbb{R}^*$ by Eqs. (37).

The functions $\zeta_{\pm}$ are sectionally analytic in the complex plane cut along the imaginary axis, with $(\zeta_{\pm})^+$ and $(\zeta_{\pm})^-$ as limits on both sides of this axis. These limits satisfy the relations (27)-(30) with $c_0=1$. Comparing the right-hand sides of these equations (with $-z_0$ instead of $z_0$), we obtain for any $z_0 \in i\mathbb{R}^*$ 
\begin{eqnarray}
H\!(a,\!z_0)(\zeta_{\pm})^+\!(a,\!b,\!-z_0)&=&H\!(a,\!z_0)(\zeta_{\pm})^{-}\!(a,\!b,\!-z_0)\nonumber\\
&\pm&\! H\!(a,\!-z_0)(\zeta_{\pm})^+\!(a,\!b,\!z_0)\exp(-b/\!z_0),
\end{eqnarray}
\begin{equation}
H\!(a,\!z_0)(\zeta_{\pm})^{-}\!(a,\!b,\!-z_0)=\mp H\!(a,\!-z_0)(\zeta_{\pm})^{-}\!(a,\!b,\!z_0)\exp(-b/\!z_0). 
\end{equation}

Note that Eq. (38) could have been deduced from Eq. (37) and the Plemelj's formulae on Cauchy-type integrals.$^{[10]}$

The functions $\zeta_{\pm}$ have remarkable analytical properties, including the following one, valid over $\mathbb{C}\!\setminus\!i\mathbb{R}^{*}$ 
\begin{equation}
\frac{1}{2}[\zeta_+\!(a,b,z)\zeta_-\!(a,b,-z)+\zeta_-\!(a,b,z)\zeta_+\!(a,b,-z)]=1, 
\end{equation}
which solely derives from the definition (37). The proof is given in the Appendix. This relation generalizes, in a finite slab, the factorization relation specific to semi-infinite media (see the conclusion). It leads to the following three properties of the functions $u_{\pm}$, $v_{\pm}$ as defined by Eqs. (15)-(16) without the 0-subscripts and with $c_0=1$:
\begin{equation}
\frac{1}{2}[u_{-}\!(a,b,z)v_{+}\!(a,b,z)+u_{+}\!(a,b,z)v_{-}\!(a,b,z)]=1\;(z\neq-1,0), 
\end{equation}
\begin{eqnarray}
\frac{1}{2}[u_{-}(a,b,z)v_{+}(a,b,-z)&-&u_{+}(a,b,z)v_{-}(a,b,-z)]=\nonumber\\
&&\qquad\exp(-b/z)\;\;(z\neq0,+1)\,,
\end{eqnarray}
\begin{eqnarray}
\frac{1}{2}[v_{+}(a,b,z)v_{-}(a,b,-z)&+&v_{-}(a,b,z)v_{+}(a,b,-z)]=\nonumber\\
&&\qquad\qquad\quad T(a,z)\;\;(z\neq\pm1)\,.
\end{eqnarray}

To summarize, the $c_0=1$ case is characterized by the following four points:

\noindent
1) The $X$- and $Y$-functions are the unique solution analytic in $\mathbb{C}^*$ to the singular integral equations (8)-(9) with $c_0=1$. They are thus the classical functions of finite plane-parallel media, first introduced by Ambartsumian$^{[11]}$ and readily studied by Chandrasekhar.$^{[12]}$ These functions are connected by the relations (6)-(7) without the 0-sub\-scripts, viz. $(z\in\mathbb{C}^*)$
\begin{eqnarray}
X(a,b,z)&=&Y(a,b,-z)\exp(-b/z),\\Y(a,b,z)&=&X(a,b,-z)\exp(-b/z).
\end{eqnarray}

Introducing the so-called Sobouti's functions$^{[13]}$
\begin{eqnarray}
\xi_{X}(a,b,z)&=&\frac{a}{2}z\int_{0}^{1}X(a,b,v)\frac{dv}{v+z}\quad(z\neq-1),\\\xi_{Y}(a,b,z)&=&\frac{a}{2}z\int_{0}^{1}Y(a,b,v)\frac{dv}{v+z}\quad(z\neq-1), 
\end{eqnarray}
the $X$- and $Y$-singular equations (8)-(9) read
\begin{eqnarray}
T(a,z)X(a,b,z)&=&1\!-\xi_{X}(a,b,\!-z)\!-\xi_{Y}(a,b,z)\exp(-b/z),\\
T(a,z)Y(a,b,z)&=&[1\!-\!\xi_{X}(a,b,z)]\!\exp(-b/z)\!-\xi_{Y}(a,b,\!-z), 
\end{eqnarray}
for any $z$ in $\mathbb{C}\setminus\lbrace0,\pm1\rbrace$.

\noindent
2) The functions $X$, $Y$, $\xi_X$ and $\xi_Y$ can be expressed in terms of the functions $u_{\pm}$ and $v_{\pm}$ by means of Eqs. (32)-(35), removing the subscripts 0 and putting $c_0=1$ in Eq. (34), which is then valid at $z=0$. Hence
\begin{eqnarray}
X(a,b,z)&=&\frac{1}{2}(u_{+}+u_{-})(a,b,z)\quad(z\in\mathbb{C}^*),\\
Y(a,b,z)&=&-\frac{1}{2}(u_{+}-u_{-})(a,b,z)\quad(z\in\mathbb{C}^*),
\end{eqnarray}
\begin{equation}
1\!-\frac{a}{2}z\!\int_{0}^{1}\!X(a,b,v)\frac{dv}{v\!+\!z}=\frac{1}{2}(v_+\! + v_-)(a,b,z)\;\; (z\!\in\!\mathbb{C}\!\setminus\!\lbrace-1\rbrace), 
\end{equation}
\begin{equation}
\frac{a}{2}z\!\int_{0}^{1}\!Y(a,b,v)\frac{dv}{v\!+\!z}=\frac{1}{2}(v_+\! -v_-)(a,b,z)\;\;(z\in\mathbb{C}\setminus\lbrace-1\rbrace).
\end{equation}

\noindent
3) The functions $u_{\pm}$ and $v_{\pm}$ follow from $\zeta_{\pm}$ using Eqs. (25)-(26) on $\mathbb{C}\setminus i\mathbb{R}$ and Eqs. (27)-(30) on $i\mathbb{R}^*$. These functions satisfy the general equations (17)-(19), which are equivalent to Eqs. (44)-(49). When $c_0 = 1$, they also verify the properties (41)-(43) resulting from the relation (40) between $\zeta_+$ and $\zeta_-$. These three equations may be rewritten in terms of the functions $X$, $Y$, $\xi_X$ and $\xi_Y$ using Eqs. (50)-(53), viz.
\begin{equation}
X(a,b,z)[1\!-\!\xi_{X}(a,b,z)]+Y(a,b,z)\xi_{Y}(a,b,z)=1\;(z\neq0,-1),
\end{equation}
\begin{eqnarray}
X(a,b,z)\xi_{Y}(a,b,-z)&+&Y(a,b,z)[1-\xi_{X}(a,b,-z)]=\nonumber\\
&&\qquad\qquad\exp(-b/z)\;\;(z\neq0,+1), 
\end{eqnarray}
\begin{eqnarray}
[1-\xi_{X}(a,b,z)][1-\xi_{X}(a,b,-z)]&-&\xi_{Y}(a,b,z)\xi_{Y}(a,b,-z)=\nonumber \\
&&\qquad T(a,z)\;\;(z\neq \pm 1). 
\end{eqnarray}

These equations are well-known: (54)-(55) are just the regular $X$- and $Y$-equations$^{[11,12]}$ and (56) was stated half a century ago.$^{[14]}$ We have retrieved the main functional properties of the functions $X$, $Y$, $\xi_X$ and $\xi_Y$ solely from the singular $X$- and $Y$-equations (48)-(49). This approach is opposite to the classical one, with central role played by the regular $X$- and $Y$-equations (54)-(55).$^{[12,15]}$ 

\noindent
4) Finally, the functions $\zeta_{\pm}$ are calculated from their definition (37), transforming the integral over the imaginary axis by the method of residues. Two Fredholm integral equations over $[0,1]$ are derived, providing an efficient algorithm for the calculation of the functions $\zeta_{\pm}$ in $\mathbb{C}\setminus i\mathbb{R}^*$. This algorithm is described in Sec. \ref{Sec5}.

To conclude on the $c_0 = 1$ case, we note that the functions $\zeta_+$ and $\zeta_-$ are at the heart of the analytical and numerical calculation of the functions $X$, $Y$, $\xi_X$ and $\xi_Y$. Obviously, those two auxiliary functions are not the only possible steps for the calculation of the $X$- and $Y$-functions. Different choices can be found  in the literature: see Busbridge,$^{[3]}$ Mullikin et al.,$^{[5,7]}$ Schultis,$^{[16]}$ Domke,$^{[17]}$ Das,$^{[18]}$ and Lahoz.$^{[19]}$ In these papers, the calculation of the $X$- and $Y$-functions amounts to the numerical solution of two Fredholm integral equations with regular kernels. The advantage of every proposed algorithm lies in the level of simplicity of the investigated integral equations, and specially on the level of regularity of their solution. Our own scheme, described in Sec. \ref{Sec5}, is based on the calculation of two functions $\zeta_+$ and $\zeta_-$ which are regular--that is defined, continuous and differentiable--everywhere on the real axis, including at 0 (contrary to $X$). Moreover, they are very smooth on the domain $[0, 1]$ of the integral equations they satisfy, whatever the values of the albedo $a \in ]0,1[$ and optical thickness $b\in ]0,+\infty[$ are. Finally they satisfy the very simple relation (40), which indicates that they are of some importance from an analytical point of view.

A different choice of auxiliary functions was made by Mullikin$^{[5]}$ forty years ago, leading up to five-digit tables of the $X$- and $Y$-functions.$^{[7]}$ Owing to the importance of his work, we have clarified, at the end of Sec. \ref{Sec5}, the link between the functions introduced by Mullikin and ours.
	  
\section{\label{Sec4}BACK TO THE GENERAL CASE}
The solution to problem (24) for any $c_0$ can be deduced from the solution $\zeta_{\pm}$ for $c_0=1$ using the following relations: 
\begin{eqnarray}
\zeta_{0,\pm}(a,b,z)&=&{\rm Y}[\Re(z)]H(a,z)c_{0}(-z)\nonumber\\&+&\frac{1}{2}\zeta_{+}(a,b,z)[\eta_{0,-}(a,b,-z)\pm \eta_{0,-}(a,b,z)]\nonumber\\&+&\frac{1}{2}\zeta_{-}(a,b,z)[\eta_{0,+}(a,b,-z)\mp \eta_{0,+}(a,b,z)], 
\end{eqnarray}
which are valid over $\mathbb{C}\setminus i \mathbb{R}$. The functions $\eta_{0,\pm}$ are defined in the whole complex plane by 
\begin{eqnarray}
\eta_{0,\pm}(a,b,z)&=&\frac{1}{2}c_{\mp}(a,b,+0)\nonumber \\
+\frac{z}{2i\pi}\!\int_{-i\infty}^{+i\infty}\!\!&H&\!\!(a,1/z')(\zeta_{\pm})^{-}\!\!(a,b,-1/z')c_{0}(-1/z')\frac{dz'}{1\!+\!zz'}. 
\end{eqnarray}
The integral is a principal value at infinity, which can be transformed into an integral over $[0,1]$ using the residue theorem. It is also a Cauchy principal value on the $i\mathbb{R}$-axis, so that the functions $\eta_{0,\pm}$ are sectionally analytic in the complex plane cut along the imaginary axis.

The proof of Eqs. (57) is not given here. It is based on the fact that their right-hand sides satisfy Eq. (24), whose unique solution is $\zeta_{0,\pm}$.

There is necessarily a link between the functions $u_{0,\pm}$ and $u_{\pm}$ on the one hand, and between the functions $v_{0,\pm}$ and $v_{\pm}$ on the other hand. This is a consequence of the relation (57) connecting the functions $\zeta_{0,\pm}$ and $\zeta_{\pm}$. It appears when substituting the expression (57) for $\zeta_{0,\pm}$ into Eqs. (25)-(26), which yields  
\begin{eqnarray}
u_{0,\pm}(a,b,z)=&\frac{1}{2}u_+(a,b,z)[\eta_{0,-}(a,b,z)\pm \eta_{0,-}(a,b,-z)]\nonumber\\+&\frac{1}{2}u_{-}(a,b,z)[\eta_{0,+}(a,b,z)\mp \eta_{0,+}(a,b,-z)], 
\end{eqnarray}
\begin{eqnarray}
v_{0,\pm}(a,b,z)=&{\rm Y}[\Re(z)]c_0(-z)\pm{\rm Y}[-\Re(z)]c_{0}(z)\exp(b/z)\nonumber\\+&\frac{1}{2}v_{+}(a,b,z)[\eta_{0,-}(a,b,-z)\pm \eta_{0,-}(a,b,z)]\nonumber\\+&\frac{1}{2}v_{-}(a,b,z)[\eta_{0,+}(a,b,-z)\mp \eta_{0,+}(a,b,z)]. 
\end{eqnarray}

Replacing the functions $u_{0,\pm}$ and $v_{0,\pm}$ by their above expressions in Eqs. (32)-(35), we derive the following solution to Eq. (1): for any $z\in\mathbb{C}^*$
\begin{equation}
X_{0}(a,\!b,\!z)=\frac{1}{2}[u_{-}(a,\!b,\!z)\eta_{0,+}(a,\!b,\!z)+u_+(a,\!b,\!z)\eta_{0,-}(a,\!b,\!z)],
\end{equation}
\begin{equation}
Y_{0}(a,\!b,\!z)=\frac{1}{2}[u_-(a,\!b,\!z)\eta_{0,+}(a,\!b,\!-z)-u_+(a,\!b,\!z)\eta_{0,-}(a,\!b,\!-z)
],  
\end{equation}
and for any $z\in \mathbb{C}\setminus\lbrace-1,0\rbrace$
\begin{eqnarray}
\lefteqn{c_{0}(-z)-\frac{a}{2}z\int_{0}^{1}X_{0}(a,b,v)\frac{dv}{v+z}={\rm Y}[\Re(z)]c_{0}(-z)}\nonumber\\&&\quad+\frac{1}{2}[v_{-}(a,b,z)\eta_{0,+}(a,b,-z)+v_{+}(a,b,z)\eta_{0,-}(a,b,-z)],
\end{eqnarray}
\begin{eqnarray}
\lefteqn{\frac{a}{2}z\int_{0}^{1}Y_{0}(a,b,v)\frac{dv}{v+z}={\rm Y}[-\Re(z)]c_{0}(z)\exp(b/z)}\nonumber\\&&\qquad-\frac{1}{2}[v
_{-}(a,b,z)\eta_{0,+}(a,b,z)-v_{+}(a,b,z)\eta_{0,-}(a,b,z)].
\end{eqnarray}

Note that Eqs. (60), (63) and (64) are still valid on $i\mathbb{R}^*$, provided that the Heaviside function $\rm Y$ is extended by 1/2 at 0.

The above expressions can be confirmed in two particular cases, viz. $b\to+\infty$ and $c_0 = 1$. When $b$ goes to infinity and the $z$-variable covers the right complex half-plane, Eq. (1) is reduced to an integral equation of the form (5), whose solution is given by (II, 50) if analyticity is required in the right complex half-plane. This solution is retrieved by letting $b\to+\infty$ in the expression (61) for $X_0(a,b,z)$. We have indeed, using evident notation, $\zeta_{\pm}(a,+\infty,z) = 1$ $\forall z\in\mathbb{C}$, $u_{\pm}(a,+\infty,z) = H(a,z)$ for $\Re(z)>0$ (these functions are not defined in the left complex half-plane), $v_{\pm}(a,+\infty,z)=1/H(a,z)$ $\forall z\in \mathbb{C}\setminus\lbrace-1\rbrace$, and
\begin{eqnarray}
\lefteqn{\eta_{0,+}(a,+\infty,z)=\eta_{0,-}(a,+\infty,z)}\nonumber\\&&\qquad\quad=\frac{1}{2}c_{0}(+0)+\frac{z}{2i\pi}\!\int_{-i\infty}^{+i\infty}\!H(a,1/z')c_{0}(-1/z')\frac{dz'}{1\!+\!zz'} 
\end{eqnarray}
for any $z\in\mathbb{C}$.

Entering these expressions into the general solution (61)-(64) leads, as $b\to+\infty$, to 
\begin{eqnarray}
X_{0}(a,+\infty,z)&=&H(a,z)\left\lbrack\frac{1}{2}c_{0}(+0)+\frac{z}{2i\pi}\!\!\int_{-i\infty}^{+i\infty}\!\!\!H(a,\!1/\!z')\right.\nonumber\\&&\left.\qquad\qquad\quad\times c_{0}(-1\!/\!z')\frac{dz'}{1\!\!+\!\!zz'}\right\rbrack\;\;[\Re(z)\!>\!0],\\ 
Y_{0}(a,+\infty,z)&=&0\quad[\Re(z)>0], 
\end{eqnarray}
\begin{eqnarray}
\lefteqn{c_{0}(-z)-\frac{a}{2}z\int_{0}^{1}X_{0}(a,+\infty,v)\frac{dv}{v+z}={\rm Y}[\Re(z)]c_{0}(-z)}\nonumber\\&&\qquad\qquad\quad+\frac{1}{H(a,z)}\,\eta_{0,+}(a,+\infty, -z)\;\;(z\in\mathbb{C}\setminus \lbrace-1,0\rbrace),      
\end{eqnarray}
\begin{equation}
\frac{a}{2}z\int_{0}^{1}Y_{0}(a,+\infty,v)\frac{dv}{v+z}=0\quad(z\in\mathbb{C}\!\setminus\!\lbrace-1\rbrace). 
\end{equation}

Equation (66) coincides with Eq. (II, 50) in the right complex half-plane, and Eq. (68) is a generalization of the classical $H$-equation (II, 31) to any $c_0(z)$. The $H$-equation is retrieved when $c_0 = 1$ since $\eta_+(a,+\infty,z)=1-{\rm Y}[-\Re(z)]H(a,-z)$.

Another interesting particular case is the one we dealt with in Sec. \ref{Sec3}, $c_0 = 1$. The functions $\eta_{\pm}$ can easily be calculated from their definition (58) with $c_0 = 1$, applying the method of residues to an obvious contour in the right complex half-plane. We find the expression
\begin{equation}
\eta_{\pm}(a,b,z)=1-{\rm Y}[-\Re(z)]H(a,-z)\zeta_{\pm}(a,b,z), 
\end{equation}
which holds on the $i\mathbb{R}$-axis adopting ${\rm Y}(0)=1/2$. Entering this expression into Eqs. (61)-(64) with $c_0=1$, we do find again the solution (50)-(53).

Coming back to the general case, we note that the definition (58) of the functions $\eta_{0,\pm}$ has to be modified for numerical purposes. The integral along the imaginary axis can be transformed into an integral over $[0,1]$ using the residue theorem. We postpone to a forthcoming paper developing the numerical solution of problem (1) for any free term $c_0$, with applications to $c_0(z) = \exp(-\tau/z)$ $(0\leq\tau\leq b)$. The main thing is that we know how to calculate the functions $\eta_{0,\pm}$ for any given $c_0(z)$, as soon as we are acquainted with the functions $\zeta_{\pm}$ over $[0,1]$. The latter functions thus yield the solution to Eq. (1) not only for the particular free term $c_0 = 1$, but for any of them as well.
       
\section{\label{Sec5} CALCULATION OF THE $\zeta_{\pm}$-FUNCTIONS}

Since the functions $\zeta_+$ and $\zeta_-$ are basic to the resolution of singular integral equations of the form (1), it is important to derive from their definition (37) an algorithm to compute them. As a matter of fact, the integral over the $i\mathbb{R}$-axis in (37) may be transformed into an integral over $[0, 1]$ using the residue theorem. We derive two uncoupled Fredholm integral equations over the range $[0,1]$, which have a unique solution and allow the calculation of the $\zeta_{\pm}$-functions outside $[0,1]$.

Replace $H(a,-1/z')$ by $1/[T(a,1/z')H(a,1/z')]$ in the integrand of Eq. (37) and integrate the function
\[
z'\to \frac{1}{T(a,1/z')}\frac{\exp(-bz')}{H^{2}(a,1/z')}\zeta_{\pm}(a,b,1/z')\frac{1}{1+zz'} 
\]
along a classical contour in the left complex half-plane (see details in Ref. [20]). One gets the following expressions of $\zeta_{\pm}(a,b,z)$ in $\mathbb{C}\setminus i\mathbb{R}$:

\noindent
- if $z\in\mathbb{C} \setminus\lbrace]-1,0[\cup\lbrace-1/k\rbrace\rbrace$
\begin{eqnarray}
\zeta_{\pm}(a,b,z)&=&1\pm M_{\pm}(a,b)\frac{2kz}{1+kz}\nonumber\\&\mp&{\rm Y}[-\Re(z)]\frac{1}{T(a,z)}\frac{\exp(b/z)}{H^{2}(a,-z)}\zeta_{\pm}(a,b,-z)\nonumber\\&\pm&\frac{a}{2}z\int_{0}^{1}(g/H^{2})(a,v)\exp(-b/v)\zeta_{\pm}(a,b,v)\frac{dv}{v+z},
\end{eqnarray}
- if $z=u \in]-1,0[$
\begin{eqnarray}
\zeta_{\pm}(a,b,u)&=&1\pm M_{\pm}(a,b)\frac{2ku}{1+ku}\nonumber\\&\mp&(gT/H^{2})(a,-u)\exp(b/u)\zeta_{\pm}(a,b,-u)\nonumber\\&\pm &\frac{a}{2}u\overset{1}{\underset{0}{f}}(g/H^2)(a,v)\exp(-b/v)\zeta_{\pm}(a,b,v)\frac{dv}{v+u},
\end{eqnarray}
where the symbol $f$ denotes the Cauchy principal value of an integral,

\noindent
- at $z=-1$
\begin{eqnarray}
\zeta_{\pm}(a,b,-1)&=&1\mp M_{\pm}(a,b)\frac{2k}{1-k}\nonumber\\&\mp& \frac{a}{2}\int_{0}^{1}(g/H^2)(a,v)\exp(-b/v)\zeta_{\pm}(a,b,v)\frac{dv}{v\!-\!1},
\end{eqnarray}
	   
\noindent
- at $z=0$
\begin{equation}
\zeta_{\pm}(a,b,0)=1,
\end{equation}

\noindent
- and at $z=-1/k$
\begin{eqnarray}
\zeta_{\pm}(a,b,-1/k)&=&1\mp\frac{2}{k}q(a,b)\zeta_{\pm}'(a,b,+1/k)\nonumber\\&\pm&2M_{\pm}(a,b)[S(a,k)-kb+\frac{2}{k}\frac{H'(a,+1/k)}{H(a,+1/k)}]\nonumber\\&\pm&\!\frac{a}{2}\!\int_{0}^{1}\!\!(g/\!H^{2})(a,\!v)\exp(-b/\!v)\zeta_{\pm}(a,\!b,\!v)\frac{dv}{1\!\!-\!kv}. 
\end{eqnarray}

We have introduced the coefficient $k=k(a)$, as the unique root in $]0,1[$ of the transcendental equation $T(a,1/z)=0$, i.e.,
\begin{equation}
T(a,1/k)=1-\frac{a}{2}\frac{1}{k}\ln\frac{1+k}{1-k}=0\quad(0<k<1).
\end{equation}

The derived coefficients $R(a,k)$, $S(a,k)$, $q(a,b)$ and $M_{\pm}(a,b)$ are defined by
\begin{equation}
R(a,k)=\frac{1-k^{2}}{k^{2}+a-1}, 
\end{equation}
\begin{equation}
S(a,k)=1-\frac{ak^2}{(k^2+a-1)(1-k^2)}, 
\end{equation}
\begin{equation}
q(a,b)=\frac{1}{2}\frac{R(a,k)}{H^{2}(a,+1/k)}\exp(-kb), 
\end{equation}
\begin{equation}
M_{\pm}(a,b)=q(a,b)\zeta_{\pm}(a,b,+1/k). 
\end{equation}

In the integrals of Eqs. (71)-(75), the notation $(g/H^2)(a,v)$ denotes the ratio $g(a,v)/H^2(a,v)$ with
\begin{equation}
g(a,v)=\frac{1}{T^2(a,v)+(\pi\frac{a}{2}v)^2}\quad(0\leq v<+1).
\end{equation}
Furthermore, we note that the derivatives at $z =+1/k$ of the functions $z\to H(a,z)$ and $z\to \zeta_{\pm}(a,b,z)$ appear in the right-hand side of Eq. (75). They can be deduced from the $H$-equation (II, 31) and Eq. (71) respectively.

Equation (71) consists of two Fredholm integral equations over $[0,1]$ for the functions $z\to \zeta_{\pm}(a,b,z)$. A difficulty arises since the coefficients $M_{\pm}(a,b)$ contain the unknown values of $\zeta_{\pm}$ at $+1/k$. It is overcome by putting $z=+1/k$ into Eq. (71) in view of calculating $M_{\pm}(a,b)$ as
\begin{eqnarray}
\lefteqn{M_{\pm}(a,b)=}\nonumber\\
&&\!\frac{q(a,b)}{1\!\mp\!q(a,b)}[1\!\pm\!\frac{a}{2}\!\int_{0}^{1}\!\!(g/H^{2})(a,v)\exp(-b/v)\zeta_{\pm}(a,b,v)\frac {dv}{1\!\!+\!kv}], 
\end{eqnarray}
which we substitute into the right-hand side of Eq. (71) with $\Re(z)>0$. This latter equation then reads in the right complex half-plane as
\begin{eqnarray}
\zeta_{\pm }(a,b,z)&=&1\pm\frac{q(a,b)}{1\mp q(a,b)}\frac{2kz}{1+kz}\nonumber\\&\pm&\frac{a}{2}\int_{0}^{1}(g/H^{2})(a,v)\exp(-b/v)\zeta_{\pm}(a,b,v)\nonumber\\&&\qquad\qquad\times\![\frac{z}{v\!+\!z}\pm \frac{q(a,b)}{1\!\mp\!q(a,b)}\frac{2kz}{1\!+\!kz}\frac{1}{1\!+\!kv}]dv.
\end{eqnarray}

The functions $\zeta_{\pm}$ are calculated by first solving numerically these two Fredholm integral equations over $[0,1]$. Once they are known in the range $[0,1]$, their values in the remainder of the complex plane are deduced from Eqs. (71)-(75) with $M_{\pm}(a,b)$ as given by Eq. (82). Although the free term and kernel of the integral Eqs. (83) are complicated, their solution can be reached easily and accurately: we computed them with ten decimal places in Ref. [20], where the functions $\zeta_{\pm}$ were introduced for the first time.$^{[21]}$ This is due to the smoothness of these functions over the $[0,1]$-range: for $b>0$, they have a finite derivative everywhere, unlike the functions $H, X$ and $\xi_X$ which have an infinite derivative at $z=0$.

Tables of the functions $\zeta_+$ and $\zeta_-$ will be published in conclusion to their detailed study on both analytical and numerical grounds. New tables of the functions $X, Y, \xi_X$ and $\xi_Y$ will also be given. The algorithm to be used is an improved version of the one described in this section. It is based on the resolution of two Fredholm integral equations much simpler than (83), since they reproduce Eqs. (83) with $q(a, b)=0$.

\noindent
{\it Link with Mullikin's algorithm for the calculation of the $X$- and $Y$-functions}

Mullikin's algorithm is based on analytic continuation techniques which lead to the solution of four Fredholm integral equations.$^{[5,7]}$ Constants are also involved, which must be computed from the solution to these four equations. Mullikin's equations are found again by letting $z\to+\infty$ in Eqs. (71), which yields
\begin{eqnarray}
\zeta_{\pm}(a,b,+\infty)&=&1\pm2M_{\pm}(a,b)\nonumber\\&\pm&\frac{a}{2}\int_{0}^{1}(g/H^{2})(a,v)\exp(-b/v)\zeta_{\pm}(a,b,v)dv, 
\end{eqnarray}
then by multiplying these equations by $kz/(1+kz)$ and subtracting the resulting equations from Eqs. (71). The new unknown functions
\begin{equation}
f(z)\pm g(z)=1\pm \frac{\zeta_{\mp}(a,b,z)}{H(a,z)}\exp(-b/z) 
\end{equation}
satisfy
\begin{eqnarray}
f(z)\pm g(z)&=&1\pm\exp(-b/z)\nonumber\\&\pm&\frac{\exp(-b/z)}{H(a,z)}[\zeta_{\mp}(a,b,+\infty)-\frac{1}{\sqrt{1-a}}]\frac{kz}{1+kz}\nonumber\\&\mp&\frac{a}{2}\frac{\exp(-b/z)}{H(a,z)}\frac{z}{1+kz}\int_{0}^{1}(g/H)(a,v)(1-kv)\nonumber\\&&\qquad\qquad\qquad\qquad\qquad\times[f(v)\pm g(v)]\frac{dv}{v+z} 
\end{eqnarray}
for $\Re(z)>0$. On the other hand, we have
\begin{equation}
\zeta_{\mp}(a,b,+\infty)=\frac{1}{\sqrt{1-a}}[1-\frac{a}{2}(\alpha_{0}\pm \beta_{0})(a,b)], 
\end{equation}
\begin{equation}
N (a,-z)=\frac{k}{\sqrt{1-a}}\frac{1}{H(a,z)}\frac{1}{1+kz}\quad(z\in\mathbb{C}\setminus[-1,0[),
\end{equation}
\begin{equation}
\frac{1}{N(a,v)}=\frac{1}{k}\sqrt{1-a}(\gamma/H)(a,v)(1-kv)\quad (v\in[0,1[). 
\end{equation}

Equations (87) follow from Eqs. (52)-(53) by letting $z\to+\infty$; the coefficients $\alpha_0$ and $\beta_0$ are the zero-order moments of the $X$- and $Y$-functions, namely
\begin{equation}
\alpha_{0}(a,b)=\int_{0}^{1}X(a,b,v)dv,\quad \beta_{0}(a,b)=\int_{0}^{1}Y(a,b,v)dv.
\end{equation}
The other two equations are deduced from Eqs. (II, 25) and (II, 20) respectively, with $\gamma=g^{1/2}$.

Equations (86) then become
\begin{eqnarray}
f(z)\pm g(z)&=&1\pm \exp(-b/z)\nonumber\\&\mp& \frac{a}{2}(\alpha_{0}\pm \beta_{0})(a,b)z\,{\exp(-b/z)}{N(a,-z)}\nonumber\\&\mp&\exp(-b/z)N(a,-z)\frac{a}{2}z\int_{0}^{1}(\gamma/N)(a,v)\nonumber\\&&\qquad\qquad\qquad\qquad\qquad\times[f(v)\pm g(v)]\frac{dv}{v+z}. 
\end{eqnarray}
This is equation (4.22) in Ref. [5], whose constants $C$ and $D$ are:
\begin{equation}
C=-\frac{a}{2}(\alpha_0+\beta_0)(a,b),\quad D=\frac{a}{2}(\alpha_0-\beta_0)(a,b). 
\end{equation}

Mullikin's functions $f$ and $g$ are thus connected to our $\zeta_{\pm}$-functions by Eqs. (85). From the very fact that the functions $\zeta_{\pm}$ are regular over the range $[-1,+1]$ (which will be proved in a forthcoming article), we deduce that the functions $f$ and $g$ are also regular in the domain $[0,1]$ of the integral equations they satisfy. This is true in particular on the right-hand side of 0. However, these two functions, contrary to $\zeta_+$ and $\zeta_-$, diverge at the ends of the range $[-1,0]$ where their evaluation is required for computing the $X$- and $Y$-functions over $[0,1]$. This behavior generates a loss of accuracy in the values of $X(a,b,u)$ and $Y(a,b,u)$ as $u$ tends to $0^+$ or $1^-$.

\section{\label{Sec6}CONCLUSION}
We solved the singular integral equations of the form (1) by first treating the case $c_0 = 1$, and then the general case. For $c_0 = 1$, the unique solution to (1) analytic in $\mathbb{C}^*$ can be written in terms of two auxiliary functions $\zeta_+$ and $\zeta_-$ which satisfy the relation (40). These functions are calculated by solving two uncoupled Fredholm integral equations over $[0,1]$. The calculation of the functions $X$, $Y$, $\xi_X$ and $\xi_Y$ follows immediately from that of the $\zeta_{\pm}$-functions: see Eqs. (50)-(53), where $u_{\pm}$ and $v_{\pm}$ are given by Eqs. (25)-(30) without the 0-subscripts. Incidently, the classical relations for the functions $X$, $Y$, $\xi_X$ and $\xi_Y$ were retrieved. 

We have then proved that the solution to Eq. (1) for any free term $c_0$ can be reduced to the one to the same equation with $c_0 = 1$. This fact undoubtedly gives weight to the $\zeta_{\pm}$-functions. The transition from the general case to the $c_0 = 1$ case is provided by the relation (57), which allows writing the general solution $X_0$ to problem (1) and related functions in the form (61)-(64). The functions $\eta_{0,\pm}$ appearing in this solution are defined by Eq. (58), which shows that they only depend on the $\zeta_{\pm}$-functions for a given $c_0$.

Our presentation is thus focused on the functions $\zeta_+$ and $\zeta_-$, the computation of which is treated in Sec. \ref{Sec5}. We could have attempted to write the general solution $X_0$ to Eq. (1) in terms of the classical $X$- and $Y$-functions. Actually, this is just what we did in a first (unpublished) version of this article. We got the formula
\begin{eqnarray}
X_{0}(a,b,z)&=&X(a,b,z)\left\lbrace\frac{1}{2}c_{0}(+0)\right.\nonumber\\&&\left.\qquad\quad+\frac{z}{2i\pi}\!\int_{-i\infty}^{+i\infty}\!X(a,b,1\!/\!z')c_{0}(-1\!/\!z')\frac{dz'}{1\!+\!zz'}\right\rbrace\nonumber\\&-&\!Y\!(a,b,z)\frac{z}{2i\pi}\!\int_{-i\infty}^{+i\infty}\!Y(a,b,1\!/\!z')c_{0}(-1\!/\!z')\frac{dz'}{1\!+\!zz'},
\end{eqnarray}
which is valid in $\mathbb{C}^*$. It yields again the solution (II, 50) in a half-space since $X(a,+\infty,z)=H(a,z)$ and $Y(a,+\infty,z)=0$ for $\Re(z)>0$. We have also $X_0 = X$ for $c_0=1$, since
\begin{eqnarray}
\frac{z}{2i\pi}\int_ {-i\infty}^{+i\infty}X(a,b,1/z')\frac{dz'}{1+zz'}&=&\frac{1}{2}-\!{\rm Y}[-\Re(z)]X(a,b,-z),\\ 
\frac{z}{2i\pi}\int_{-i\infty}^{+i\infty}Y(a,b,1/z')\frac{dz'}{1+zz'}&=&-{\rm Y}[-\Re(z)]Y(a,b,-z), 
\end{eqnarray}
from the residue theorem. However, we stress that neither the $X$- and $Y$-functions, nor the $\xi_ X$- and $\xi_Y$-functions are proper intermediate steps to solve the radiative transfer equation within a slab. This important question will be discussed in a forthcoming paper while solving the integral form of the transfer equation. Anticipating this discussion, we conclude that the expression (93) of the solution to problem (1) is more elegant but less judicious than the expression (61) based on the $\zeta_{\pm}$-functions.

Finally, we note that there are several possible generalizations to the $H$-function in a finite slab. The classical approach consists in introducing the couple $(X,Y)$, which yields $(H,0)$ in the right complex half-plane when the optical thickness $b$ goes to infinity. Another possibility is provided by the functions
\begin{equation}
H_{+}(a,b,z)=H(a,z)\zeta_{+}(a,b,z),
\end{equation}
\begin{equation}
H_{-}(a,b,z)=H(a,z)\zeta_{-}(a,b,z),
\end{equation} 
which verify from Eq. (40)
\begin{equation}
T(a,\!z)\frac{1}{2}[H_{+}(a,\!b,\!z)H_{-}(a,\!b,\!-z)+H_{-}(a,\!b,\!z)H_{+}(a,\!b,\!-z)] =1. 
\end{equation}

These two functions coincide everywhere with the $H$-function as $b\to+\infty$ [since $\zeta_{\pm}(a,+\infty,z)= 1$], and we retrieve the factorization relation $T(a,z)H(a,z)H(a,-z)=1$ by letting $b\to+\infty$ in Eq. (98).

\section*{APPENDIX: PROOF OF EQ. (40)}

Equation (40) can been proved by first calculating the product $\zeta_+(a,b,z)\zeta_-(a,b,-z)$ from the definition (37) of the functions $\zeta_{\pm}$. Observing that
\[
\frac{1}{(1+zz')(1-zz")}=\frac{1}{z'+z"}[\frac{z'}{1+zz'}+\frac{z"}{1-zz"}],
\]
the product of the two integrals along the $i\mathbb{R}$-axis is transformed into a sum of two repeated integrals. This sum is calculated by inverting the order of integration in one of the repeated integrals (which is allowed when $z\notin i\mathbb{R}^*$), replacing then the Cauchy integrals over $i\mathbb{R}$ by their expression as given by Plemelj's formulae.$^{[10]}$ After some simplifications, one obtains
\begin{eqnarray}
\zeta_+(a,b,z)\zeta_-(a,b,-z)&=&1+ \frac{z}{2i\pi}\int_{-i\infty}^{+i\infty}\frac{H(a,-1/z')}{H(a,1/z')}\exp(-bz')\nonumber\\&&\qquad\qquad\times[(\zeta_+)^+\!\!+\!(\zeta_-)^+](a,b,1/z')dz'\nonumber\\&+&\frac{2z^3}{2i\pi}\int_{-i\infty}^{+i\infty}\frac{H(a,-1/z')}{H(a,1/z')}\exp(-bz')\nonumber\\&&\quad\times[(\zeta_+)^+\!\!\times\!(\zeta_-)^+](a,\!b,\!1/z')\frac{z'^2 dz'}{1\!-\!z^2z'^2},\nonumber
\end{eqnarray}
which shows that the product $\zeta_+(a,b,z)\zeta_-(a,b,-z)$ has the form $1+\varphi(z)$, where $\varphi$ is an odd function of $z$. Changing $z$ into $-z$, we deduce that $\zeta_+(a,b,-z)\zeta_-(a,b,z)=1+\varphi(-z)=1-\varphi(z)$. Hence the relation (40) for $z\in\mathbb{C}\setminus i\mathbb{R}^*$.

\end{document}